\documentclass{ifacconf}

\usepackage{graphicx}      
\usepackage{natbib}        

\usepackage{cancel}
\usepackage[usenames,dvipsnames]{color}

\usepackage{amsmath,color}
\usepackage{amssymb}
\usepackage{amsfonts}
\usepackage{graphicx}
\usepackage{epstopdf}

\usepackage{enumerate}
\usepackage{amsmath}
\usepackage{cancel}
\usepackage{mathrsfs}
\usepackage{mathdots}
\usepackage{euscript}
\usepackage{amscd}
\usepackage{placeins}
\usepackage{tikz}
\usetikzlibrary{snakes,arrows,shapes}
\usepackage[toc,page]{appendix}

\graphicspath{{images/}}

\theoremstyle{definition}\newtheorem{ex}[thm]{Example}

\newcommand{\bmat}{\left[ \begin{matrix}}
	\newcommand{\emat}{\end{matrix} \right]}

\newcommand{\trace} {\mbox{\rm tr}}

\newcommand{\E}{{\mathbb E}}
\newcommand{\Rbb}{\mathbb R}

\newcommand{\Cbb}{\mathbb C}
\newcommand{\Dbb}{\mathbb D}

\newcommand{\Tbb}{\mathbb T}

\newcommand{\image}{\mathrm{im}}
\newcommand{\adj}{\mathrm{adj}}
\newcommand{\diag}{\mathrm{diag}}

\newcommand{\Hfrak}{\mathfrak{H}}

\newcommand{\Cfrak}{\mathfrak{C}}

\newcommand{\Sscr}{\mathscr{S}}
\newcommand{\Lscr}{\mathscr{L}}
\newcommand{\Cscr}{\mathscr{C}}

\newcommand{\Scal}{\mathcal{S}}

\begin{document}
\begin{frontmatter}

\title{On a Parametric Spectral Estimation Problem\thanksref{footnoteinfo}} 

\thanks[footnoteinfo]{This work was supported by the China Scholarship Council (CSC) under File No. 201506230140.}

\author[First]{B. Zhu} 

\address[First]{Department of Information Engineering, University of Padova, Via Gradenigo 6/B, 35131 Padova, Italy (e-mail: zhubin@dei.unipd.it)}

\begin{abstract}                
We consider an open question posed in \cite{Zhu-Baggio-17} on the uniqueness of the solution to a parametric spectral estimation problem.
\end{abstract}


\end{frontmatter}

\section{Introduction}\label{SecIntro}

In this paper, we consider a spectral estimation problem subjected to a generalized moment constraint, a framework pioneered by Byrnes, Georgiou, and Lindquist in \cite{BGL-THREE-00,Georgiou-L-03}. The formulation of the problem can be seen as a generalization of earlier work on \emph{rational covariance extension}, cf. e.g., \cite{Kalman,Gthesis,BLGM-95,BGL-98,SIGEST-01}, and \emph{Nevanlinna--Pick interpolation}, cf. \cite{Georgiou-87-NP,BGL-01} and references therein.

A standard setup of the problem is as follows. Suppose we have a zero-mean wide-sense stationary vector signal $y(t)$ with an unknown spectral density matrix $\Phi(z)$. In order to estimate the spectrum, we perform the following steps.

\begin{enumerate}[Step 1.]
\item Feed the signal $y(t)$ into a filter bank with a transfer function
\begin{equation}\label{trans_func}
G(z)=(zI-A)^{-1}B
\end{equation}
to get an output $x(t)$. The corresponding time domain representation is just
\begin{equation}\label{filter_bank}
x(t+1)=Ax(t)+By(t).
\end{equation}
We have some extra specifications on the system matrices, which include
\begin{itemize}
\item $A\in\Cbb^{n\times n}$ is Schur stable, i.e., all its eigenvalues have moduli less than $1$;
\item $B\in\Cbb^{n\times m}$ is of full column rank with $n\geq m$;
\item The pair $(A,B)$ is \emph{reachable}.
\end{itemize}

\item Compute an estimate of the steady-state covariance matrix $\Sigma:=\E\,x(t)x(t)^*$ of the state vector $x(t)$; cf.~e.g.,~\cite{FPZ-12} for such structured covariance estimation problem. Hence we have
\begin{equation}\label{mmt_constraint}
\int G\Phi G^*=\Sigma,
\end{equation}
where the function is integrated on the unit circle $\Tbb$ against the normalized Lebesgue measure, i.e.,
$$\int F:=\int_{-\pi}^{\pi}F(e^{i\theta})\frac{d\theta}{2\pi}.$$
This simplified notation will be adopted throughout.

\item Given the estimated $\Sigma>0$, find a spectral density $\Phi$ such that the generalized moment constraint (\ref{mmt_constraint}) is satisfied.
\end{enumerate}


We must point out that existence of a bounded and coercive $\Phi$ satisfying (\ref{mmt_constraint}) is not trivial in general. Such feasibility problem was addressed in \cite{Georgiou-02}, see also e.g., \cite{FPZ-10,FPZ-12}. In this paper, we shall always assume the feasibility in the sense explained next.
Let $C(\Tbb;\Hfrak_m)$ denote the space of $m\times m$ Hermitian matrix-valued continuous functions on the unit circle and let $\Hfrak_{n}$ be the vector space of $n\times n$ Hermitian matrices.
Define the linear operator
\begin{equation}\label{map_Gamma}
\begin{split}
\Gamma:C(\Tbb;\Hfrak_m) & \to\Hfrak_n \\
\Phi & \mapsto\int G\Phi G^*.
\end{split}
\end{equation}
We shall denote the image/range of this map by $\image\,\Gamma$ for short. Then we assume that the covariance matrix $\Sigma\in\image\,\Gamma$. According to \cite[Proposition 3.1]{FPZ-12}, $\image\,\Gamma$ is a linear space with real dimension $m(2n-m)$.

Given a positive definite $\Sigma\in\image\,\Gamma$, there are in general infinitely many spectral densities that would solve (\ref{mmt_constraint}). The mainstream approach today to remedy such ill-posedness is to first introduce a prior matrix spectral density $\Psi$, which represents our guess of the ``true'' density $\Phi$. Then one tries to define an entropy-like distance index $d(\Phi,\Psi)$ between two spectral densities, and to find the ``best'' $\Phi$ by solving the constrained optimization problem
\[\underset{\Phi\in\Scal_m}{\text{minimize}}\ d(\Phi,\Psi)\text{ subject to } (\ref{mmt_constraint}),\]
where $\Scal_m$ is the family of $m\times m$ bounded and coercive spectral densities. Due to the page limit, we refer the readers to e.g., \cite{Zhu-Baggio-17} for a brief review of the literature in this direction. 


In this work however, we attempt to attack the problem in a direction different from optimization, as a continuation of the work in \cite{FPZ-10}, where a parametric family of spectral densities was introduced, and a certain map from the parameter space to the space of generalized moments was studied. 
The question whether a solution to the parametric spectral estimation problem in fact exists was essentially left open in \cite{FPZ-10} until recently, such existence result has been worked out in \cite{Zhu-Baggio-17}. In this paper, we try to approach the question of uniqueness of the solution and even well-posedness of the problem. The main tool here is the global inverse function theorem of Hadamard that is reported e.g., in \cite{Gordon_diffeo-72}. However, we do not claim to have answered such questions to a satisfactory level. Instead, we only provide a possible way to the answer.



The outline of this paper is as follows.
In Section \ref{SecProb}, we review the problem formulation and characterize the solution in a special case. A spectral factorization problem is discussed in Section \ref{SecSpecFact}, whose result will be useful for the development in Section \ref{SecGeneral}, where we present our main results.

\section{A parametric formulation and the solution in a special case}\label{SecProb}

Let us first define the set
\begin{equation}
\Lscr_+:=\{\Lambda\in\Hfrak_{n}\;:\;G^*(z)\Lambda G(z)>0,\ \forall z\in\Tbb\},
\end{equation}
which obviously contains all the Hermitian positive definite matrices, since $G(z)$ is of full column rank for any $z\in\Tbb$ which readily follows from the problem setup.
By the continuous dependence of eigenvalues on the matrix entries, one can verify that $\Lscr_+$ is an open subset of $\Hfrak_{n}$. 

For $\Lambda\in\Lscr_+$, take $W_\Lambda$ as the unique stable and minimum phase {(right)} spectral factor of $G^*\Lambda G$ \cite[Lemma 11.4.1]{FPZ-10}, i.e.,
\begin{equation}\label{W_Lambda}
G^*\Lambda G=W_\Lambda^*W_\Lambda.
\end{equation}
The spectral factor $W_{\Lambda}$ can be written as
\begin{equation}\label{W_spec_factor}
W_{\Lambda} (z) = L^{-*}B^{*}PA(zI -A)^{-1}B + L,
\end{equation}
where $P$ is the unique stabilizing solution of the Discrete-time Algebraic Riccati Equation (DARE)
\begin{equation}\label{DARE}
\Pi = A^{*}\Pi A- A^{*}\Pi B(B^{*}\Pi B)^{-1}B^{*}\Pi A+\Lambda,
\end{equation}
and $L$ is the right Cholesky factor of the positive matrix $B^{*}PB$, i.e.,
\begin{equation}\label{L_Chol_factor}
B^*PB=L^*L
\end{equation}
with $L$ being lower triangular having real and positive diagonal entries. It is worth pointing out that the DARE (\ref{DARE}) above is not a standard one, as $\Lambda\in\Lscr_+$ can be indefinite. A formal proof for the existence of a stabilizing solution can be found in the appendix of \cite[Paper A]{avventi2011spectral}.

To avoid any redundancy in the parameterization, we have to define the set $\Lscr_+^\Gamma:=\Lscr_+\cap\image\,\Gamma$. This is due to a simple geometric result. More precisely, the adjoint map of $\Gamma$ in (\ref{map_Gamma}) is given by (cf. \cite{FPZ-10})
\begin{equation}\label{Gamma_star}
\begin{split}
\Gamma^*:\Hfrak_n & \to C(\Tbb;\Hfrak_m) \\
X & \mapsto G^*XG,
\end{split}
\end{equation}
and we have the relation
\begin{equation}\label{Range_Gamma_ortho}
\left(\image\,\Gamma\right)^\perp=\ker\Gamma^*=\left\{X\in\Hfrak_n\,:\,G^*(z)XG(z)=0,\ \forall z\in\Tbb\right\}.
\end{equation}
Hence for any $\Lambda\in\Lscr_+$, we have the orthogonal decomposition
\[\Lambda=\Lambda^\Gamma+\Lambda^\perp\]
with $\Lambda^\Gamma\in\image\,\Gamma$ and $\Lambda^\perp$ in the orthogonal complement. In view of (\ref{Range_Gamma_ortho}), the part $\Lambda^\perp$ does not contribute to the function value of $G^*\Lambda G$ on the unit circle, and we simply have
\[\Lscr_+^\Gamma=\Pi_{\image\,\Gamma}\Lscr_+,\]
where $\Pi_{\image\,\Gamma}$ denotes the orthogonal projection operator onto the linear space $\image\,\Gamma$.

From this point on, we shall take the prior $\Psi\in\Scal_m$ to be continuous on $\Tbb$, which would facilitate reasoning. We can now define a parametric family of spectral densities
\begin{equation}\label{family_SpecDen}
\Sscr:=\{\,\Phi_\Lambda=W_\Lambda^{-1}\Psi W_\Lambda^{-*}\,:\,\Lambda\in\Lscr_+^\Gamma\,\}.
\end{equation}
We have the map 
\[\Lambda\mapsto W_\Lambda\mapsto W_\Lambda^{-1}\Psi W_\Lambda^{-*}\]
from the parameter $\Lambda\in\Lscr_+^\Gamma$ to the density function $\Phi_\Lambda$. 

\begin{rem}
In the scalar case, the form of spectral densities in the family (\ref{family_SpecDen}) reduces to
$$\Phi_\Lambda = \frac{\Psi}{G^*\Lambda G}\,,$$
which is precisely the solution (4.3) in \cite{Georgiou-L-03} of a constrained optimization
problem in terms of the Lagrange multiplier $\Lambda$. An alternative matricial parametrization has been proposed and studied in \cite{Georgiou-06}.
\end{rem}

Our problem is formulated as follows.

\begin{prob}\label{spec_estimation}
	Given the filter bank $G(z)$ in (\ref{trans_func}), the prior $\Psi\in\Scal_m$ continuous, and a positive definite matrix $\Sigma\in\image\,\Gamma$, find a spectral density in the parametric family $\Sscr$ defined in (\ref{family_SpecDen}) such that
	\begin{equation}\label{constraint}
	\int G\Phi_\Lambda G^*=\Sigma.
	\end{equation}
\end{prob}

The above problem has an equivalent formulation. Define $\image_+\Gamma:=\image\,\Gamma\cap\Hfrak_{+,n}$ where $\Hfrak_{+,n}$ is the open set of $n\times n$ Hermitian positive definite matrices. Consider the map
\begin{equation}\label{map_omega}
\begin{split}
\omega\colon\Lscr_+^\Gamma & \to\image_+\Gamma \\
\Lambda & \mapsto\int G\Phi_\Lambda G^*.
\end{split}
\end{equation}
Then Problem \ref{spec_estimation} is asking: what is the preimage of $\Sigma\in\image_+\Gamma$ under the map $\omega\,$? As shown in \cite{Zhu-Baggio-17}, this is a continuous \emph{surjective} map between open subsets of the linear space $\image\,\Gamma$, and thus a solution to Problem \ref{spec_estimation} always exists. The question now is whether the solution is unique. We show next that uniqueness is indeed true if the prior $\Psi$ has a special structure.

\subsection{Well-posedness given a scalar prior}

In the case of a scalar prior, in which we take $\Psi(z)=\psi(z)I_m$, where the scalar-valued function $\psi(z)\in\Sscr_1$ is continuous, the map $\omega$ would reduce to
\begin{equation}\label{map_omega_tilde}
\begin{split}
\tilde{\omega}\colon\Lscr_+^\Gamma & \to\image_+\Gamma \\
\Lambda & \mapsto\int \psi G(G^*\Lambda G)^{-1}G^*,
\end{split}
\end{equation}
and the family of spectral densities becomes
\begin{equation}\label{family_SpecDen_scalar}
\tilde{\Sscr}:=\{\,\Phi_\Lambda=\psi(G^*\Lambda G)^{-1}\,:\,\Lambda\in\Lscr_+^\Gamma\,\}.
\end{equation}


According to \cite{FPZ-10}, solution to Problem \ref{spec_estimation} under a scalar prior exists and is unique. We shall next show that given a continuous prior $\psi$, the map $\tilde{\omega}$ is a $C^1$ \emph{diffeomorphism}\footnote{The word ``diffeomorphism'' in the sequel should always be understood in the $C^1$ sense. Hence the attributive $C^1$ will be omitted.}between $\Lscr_+^\Gamma$ and $\image_+\Gamma$, which in particular, means that the solution $\Lambda$ depends continuously on the covariance data $\Sigma$, and thus the problem is well-posed in the sense of Hadamard. The proof is an application of the global inverse function theorem of Hadamard that appears e.g., in \cite{Gordon_diffeo-72}.

\begin{thm}[Hadamard]\label{Hadamard}
	Let $M_1$ and $M_2$ be connected, oriented, boundary-less $n$-dimensional manifolds of class $C^1$, and suppose that $M_2$ is simply connected. Then a $C^1$ map $f\colon M_1 \to M_2$ is a diffeomorphism if and only if $f$ is proper and the Jacobian determinant of $f$ never vanishes.
\end{thm}
Conditions on the domain and codomain of $\tilde{\omega}$ can be verified easily. In fact, the set $\Lscr_+^\Gamma=\Lscr_+\cap\image\,\Gamma$ is easily seen to be open and path-connected since both $\Lscr_+$ and $\image\,\Gamma$ are such. The simple connectedness of $\image_+\Gamma$ has been reported in \cite[Proposition 1]{Zhu-Baggio-17}. The fact that $\tilde{\omega}$ is of class $C^1$ can be seen along the proof of \cite[Lemma 1]{Zhu-Baggio-17}. Moreover, properness of the more general map $\omega$ has been proven in \cite[Theorem 11.4.1]{FPZ-10}. Therefore, it is only left to check the Jacobian of $\tilde{\omega}$. The next result can be viewed as an interpretation of \cite[Theorem 11.4.2]{FPZ-10}. Here and in the sequel, we shall introduce the notation $\Phi(z;\Lambda)$ to denote a spectral density function that depends on the parameter $\Lambda$, and use it interchangeably with $\Phi_\Lambda(z)$.
\begin{prop}\label{prop_diffeo_scalar}
The Jacobian determinant of $\tilde{\omega}$ never vanishes in $\Lscr_+^\Gamma$, and hence the map $\tilde{\omega}$ is a diffeomorphism.
\end{prop}
\begin{pf}
    From \cite[Lemma 1]{Zhu-Baggio-17}, the differential of $\tilde{\omega}$ at $\Lambda\in\Lscr_+^\Gamma$ is
	\begin{equation}\label{diff_omega_tilde}
	\delta\tilde{\omega}(\Lambda;\delta\Lambda)=-\int \psi G(G^*\Lambda G)^{-1}(G^*\delta\Lambda G)(G^*\Lambda G)^{-1}G^*
	\end{equation}
	such that $\delta\Lambda\in\image\,\Gamma$. Our target is to show that
	\[\delta\tilde{\omega}(\Lambda;\delta\Lambda)=0\implies\delta\Lambda=0.\]
	To this end, first notice that the middle part of the integrand in (\ref{diff_omega_tilde}) is just the differential of the spectral density $\Phi_\Lambda=\psi(G^*\Lambda G)^{-1}$ w.r.t. $\Lambda\,$: \[\delta\Phi(z;\Lambda;\delta\Lambda):=-\psi (G^*\Lambda G)^{-1}(G^*\delta\Lambda G)(G^*\Lambda G)^{-1}.\] 
	Then the condition $\delta\tilde{\omega}(\Lambda;\delta\Lambda)=0$ means that
	\[\delta\Phi(z;\Lambda;\delta\Lambda)\in\ker\Gamma=(\image\,\Gamma^*)^\perp,\]
	which in view of (\ref{Gamma_star}), reads
	\[\begin{split}
	\langle G^*XG,\delta\Phi(z;\Lambda;\delta\Lambda)\rangle & =\trace\int G^*XG\,\delta\Phi(z;\Lambda;\delta\Lambda) \\
	 & =0,\quad \forall X\in\Hfrak_n.
	\end{split}\]
	In particular, following \cite[Eqns.~11.44--11.45]{FPZ-10}, choosing $X=\delta\Lambda$ would lead to
	\[G^*\delta\Lambda G\equiv0,\quad\forall z\in\Tbb,\]
	which by (\ref{Range_Gamma_ortho}), implies that $\delta\Lambda\in(\image\,\Gamma)^\perp$. Since at the same time $\delta\Lambda\in\image\,\Gamma$, it is necessary that $\delta\Lambda=0$. The rest is just an application of Theorem \ref{Hadamard}.
	
\end{pf}

\begin{rem}
The unique solution in $\tilde{\Sscr}$ to the spectral estimation problem has an interesting characterization in terms of an optimization problem; cf. \cite[Paper A]{avventi2011spectral} for details.
\end{rem}

A difficulty arises when one tries to extend the analysis in the previous proposition to the more general map $\omega$, as it would entail the differentiation of the spectral factor $W_\Lambda$ in (\ref{W_Lambda}) w.r.t. the parameter $\Lambda$. Such a difficulty can be bypassed by introducing a spectral factorization as will be discussed next.

\section{A diffeomorphic spectral factorization}\label{SecSpecFact}

Following the lines of \cite{avventi2011spectral}, given the stabilizing solution $P$ of the DARE (\ref{DARE}), let us introduce a change of variables by setting
\begin{equation}\label{C_factor}
C:=L^{-*}B^{*}P.
\end{equation}
Then it is not difficult to recover the relation $L=CB$ for the Cholesky factor in (\ref{L_Chol_factor}). In this way, the spectral factor (\ref{W_spec_factor}) can be rewritten as
\begin{equation}\label{W_factor_C}
\begin{split}
W_{\Lambda} (z) & = CA(zI -A)^{-1}B + CB \\
 & =zCG,
\end{split}
\end{equation}
where the second equality holds because of the identity $A(zI-A)^{-1}+I=z(zI-A)^{-1}$. In view of this, the factorization (\ref{W_Lambda}) can then be rewritten as
\begin{equation}\label{spec_fact_Lambda}
G^*\Lambda G=G^*C^*CG, \quad \forall z\in\Tbb.
\end{equation}
This relation has also been expressed in \cite[Equation 11.29]{FPZ-10}. 
In the sequel, we shall also call the $m\times n$ matrix $C$ a `` spectral factor''. 

As reported in \cite[Section A.5.5]{avventi2011spectral}, it is possible to build a \emph{homeomorphic} factorization by carefully choosing the set where the factor $C$ lives. More precisely, let the set $\Cscr_+\subset\Cbb^{m\times n}$ contain those matrices $C$ that satisfy the following two conditions
\begin{itemize}
	\item $CB$ is lower triangular with real and positive diagonal entries,
	\item $A-B(CB)^{-1}CA$ has eigenvalues strictly inside the unit circle.
\end{itemize}
Define 
the map
\begin{equation}\label{H_map}
\begin{split}
h:\,\Lscr_+^\Gamma & \to\Cscr_+ \\
\Lambda & \mapsto C\textrm{ via } (\ref{C_factor}).
\end{split}
\end{equation}
Then according to \cite[Theorem A.5.5]{avventi2011spectral},	the map $h$ of spectral factorization is a homeomorphism. We shall next strengthen this result by showing that the map $h$ is in fact a {diffeomorphism} using Theorem \ref{Hadamard}.

\subsection{Characterization of diffeomorphism}\label{SubsecDiffeo}

We are going to apply Theorem \ref{Hadamard} to the inverse of $h$
\begin{equation}\label{H_inverse}
\begin{split}
h^{-1}:\,\Cscr_+ & \to\Lscr_+^\Gamma \\
C & \mapsto \Lambda:=\Pi_{\image\,\Gamma}(C^*C).
\end{split}
\end{equation}
Those technical requirements on the domain and codomain of $h^{-1}$ can be verified without difficulty. The set $\Cscr_+$ is an open subset of the linear space 
\[\begin{aligned}
\Cfrak:=\left\{\,C\in\Cbb^{m\times n}\right.\,: & \, CB \textrm{ is lower triangular} \\
 & \textrm{with real diagonal entries}\,\},
\end{aligned}\] 
whose real dimension coincides with $\image\,\Gamma$ (cf. \cite{avventi2011spectral}). The fact that $\Cscr_+$ is also path-connected is a consequence of $h$ being a homeomorphism. Furthermore, the proof of $\Lscr_+^\Gamma$ being simply connected can be adapted easily from \cite[Proposition 1]{Zhu-Baggio-17}.

The map $h^{-1}$ is actually smooth (hence of course $C^1$) because it is a composition of the quadratic map $C\mapsto C^*C$ and the projection $\Pi_{\image\,\Gamma}$, both of which are smooth. The fact that $h^{-1}$ is proper has also been reported in \cite{avventi2011spectral}. Therefore, it remains to investigate the Jacobian of $h^{-1}$. In order to carry out explicit computation, it is necessary to choose bases for the two linear spaces $\Cfrak$ and $\image\,\Gamma$. 

Let $M:=m(2n-m)$, and let $\{\Lambda_1,\Lambda_2,\dots,\Lambda_M\}$ and $\{C_1,\dots,C_M\}$ be orthonormal bases of $\image\,\Gamma$ and $\Cfrak$, respectively. Then one can parameterize $\Lambda\in\Lscr_+^\Gamma$ and $C\in\Cscr_+$ as
\begin{equation}\label{Lambda_C_coord}
\begin{split}
\Lambda(x) & =x_1\Lambda_1+x_2\Lambda_2+\cdots+x_M\Lambda_M, \\
C(y) & =y_1C_1+y_2C_2+\cdots+y_MC_M,
\end{split}
\end{equation}
for some $x_j,y_j\in\Rbb$, $j=1,\dots,M$.
The map $h^{-1}$ can then be expressed coordinate-wisely as
\begin{equation}\label{H_inv_coordinate}
x_j=\langle\Lambda_j,C(y)^*C(y)\rangle.
\end{equation}
Then the partial derivatives can be computed as
\begin{equation}\label{partial_derivatives}
\frac{\partial x_j}{\partial y_k}=\langle\Lambda_j,C_k^*C(y)+C^*(y)C_k\rangle,
\end{equation}
which is the $(j,k)$ element of the Jacobian matrix denoted as $J_{h^{-1}}(y)$. We need some ancillary results in order to show that $h^{-1}$ has everywhere nonvanishing Jacobian.

\begin{prop}\label{prop_nonsingular}
	If $v\in\Cbb^n$ is such that $v^*G(z)=0$ for all $z\in\Tbb$, then $v=0$. 
\end{prop}
\begin{pf}
    The condition that $v^*G(z)=0$ for all $z\in\Tbb$ implies that
    \[v^*\int GG^* v=0.\]
    Under our problem setting stated in Section \ref{SecIntro}, we have $\int GG^*>0$ and thus the assertion of the proposition follows. To see the fact of positive definiteness, note first that the following expansion holds
    \begin{equation}\label{power_series_expan}
    \begin{split}
    G(z) & =(zI-A)^{-1}B \\
     & =z^{-1}\sum_{k=0}^{\infty}z^{-k}A^kB,\quad \textrm{for } |z|\geq 1,
    \end{split}
    \end{equation}
    since $A$ is stable. Then by the Parseval identity, we have
    \[\int GG^*=\sum_{k=0}^{\infty}A^kBB^*(A^*)^k=RR^*,\]
    where $R=[B,AB,\dots,A^kB,\dots]$. The above is the unique solution of the discrete-time Lyapunov equation
	\begin{equation}
	X-AXA^*=BB^*.
	\end{equation}
	Since $(A,B)$ is by assumption reachable, $R$ is of full row rank, and therefore $\int GG^*>0$.

\end{pf}



\begin{prop}\label{prop_rational_mat_eqn}
	Given $C\in\Cscr_+$, the rational matrix equation in the unknown $V\in\Cbb^{m\times n}$ 
	\begin{equation}\label{rational_eqn}
	G^*(C^*V+V^*C)G=0,\quad \forall z\in\Tbb
	\end{equation}
	has the general solution
	\begin{equation}
	V=QC
	\end{equation}
	where $Q\in\Cbb^{m\times m}$ is an arbitrary constant skew-Hermitian matrix. If one further requires $V\in\Cfrak$, then $(\ref{rational_eqn})$ has only the trivial solution $V=0$.
\end{prop}
\begin{pf}
	The equation (\ref{rational_eqn}) is equivalent to
	\begin{equation}\label{rational_eqn_equiv}
	z^{*}G^*(C^*V+V^*C)Gz=0,\quad \forall z\in\Tbb.
	\end{equation}
	Let
	\[\begin{split}
	zCG(z) & =zC(zI-A)^{-1}B \\
	 & =\frac{P_C(z)}{z^{-n}\det(zI-A)},
	\end{split}\]
	where $P_C(z):=z^{-n+1}C\,\adj(zI-A)\,B$ and $\adj(\cdot)$ denotes the adjugate matrix. Obviously, $P_C(z)$ is a matrix polynomial in the indeterminate $z^{-1}$, which is intended to conform to the engineering convention. From (\ref{power_series_expan}), we have
	\[\lim_{z\to\infty}zCG=CB=\lim_{z\to\infty}P_C(z),\]
	where the second equality holds since $\lim_{z\to\infty}z^{-n}\det(zI-A)=1$.
	Moreover, the scalar polynomial $\det P_C(z)$ has all its roots inside $\Dbb$, which can be seen from (\ref{W_factor_C}) as $zCG$ is minimum phase, i.e., admits a stable inverse. 
	
	Define similarly $P_V(z):=z^{-n+1}V\,\adj(zI-A)\,B$. Then one can reduce (\ref{rational_eqn_equiv}) to the matrix polynomial equation	
	\begin{equation}\label{poly_eqn}
	P_C^*(z)P_V(z)+P_V^*(z)P_C(z)=0,\quad \forall z\in\Tbb,
	\end{equation}
	in which we have
	\[P^*_C(0)=\left[\lim_{z\to\infty}P_C(z)\right]^*=(CB)^*\]
	nonsingular because $C\in\Cscr_+$.
	By the identity theorem for holomorphic functions, if the above equation holds on $\Tbb$, then it holds for any $z\in\Cbb$ except for $0$ (and $\infty$). Hence the restriction $z\in\Tbb$ can be removed here. Since $P_C^*$ is anti-stable and $P_C^*(0)$ nonsingular, according to (a variant of) \cite[Theorem MP1]{jevzek1986symmetric}, the general solution of (\ref{poly_eqn}) is
	\[P_V=QP_C,\]
	where $Q\in\Cbb^{m\times m}$ is an arbitrary constant skew-Hermitian matrix. This in turn implies that
	\begin{equation}\label{rational_identity}
	VG(z)=QCG(z),\quad \forall z\in\Tbb,
	\end{equation}
	which in view of Proposition \ref{prop_nonsingular}, further implies that $V=QC$.
	
	To prove the remaining part of the claim, just apply the power series expansion (\ref{power_series_expan}) to (\ref{rational_identity}), and notice that all the Fourier coefficients on the two sides of (\ref{rational_identity}) must coincide. This in particular means that
	\[VB=QCB.\]
	Since we have $C\in\Cscr_+$ and $V\in\Cfrak$ in addition, both $VB$ and $CB$ are lower triangular and the latter is invertible. Therefore $Q$ turns out to be also lower triangular and at the same time skew-Hermitian, which necessarily means that $Q$ is equal to $0$ and so is $V$.
	
\end{pf}

\begin{thm}\label{thm_diffeo_H}
	The Jacobian determinant of $h^{-1}$ never vanishes in $\Cscr_+$, and hence the map $h$ in $(\ref{H_map})$ is a diffeomorphism.
\end{thm}
\begin{pf}
Suppose $v\in\Rbb^M$ is such that $J_{h^{-1}}(y)v=0$. We need to show that $v=0$. To this end, notice from (\ref{partial_derivatives}) that equivalently we have for $j=1,2,\dots,M$,
\[\begin{split}
0 & =\sum_{k=1}^{M}v_k\langle\Lambda_j,C_k^*C(y)+C^*(y)C_k\rangle \\
 & =\langle\Lambda_j,C^*(v)C(y)+C^*(y)C(v)\rangle,
\end{split}\]
which implies that
\[C^*(v)C(y)+C^*(y)C(v)\perp\image\,\Gamma.\]
In view of (\ref{Range_Gamma_ortho}), this in turn means
\[G^*(z)\left[C^*(v)C(y)+C^*(y)C(v)\right]G(z)=0,\quad\forall z\in\Tbb.\]
By Proposition \ref{prop_rational_mat_eqn}, the only solution is $v=0$. Thus Theorem \ref{Hadamard} is applicable and this completes the proof.

\end{pf}

\section{The general map $\omega$}\label{SecGeneral}
Let us return to the map $\omega$ defined in (\ref{map_omega}). We shall use the result obtained in the previous section to attack the uniqueness conjecture posed in \cite{Zhu-Baggio-17}. 
Given the relation (\ref{W_factor_C}), the spectral density $\Phi_\Lambda$ can be reparameterized in $C$ as
\begin{equation}
\Phi_\Lambda\equiv\Phi_C:=(CG)^{-1}\Psi(CG)^{-*}.
\end{equation}
In this way, the map $\omega$ can be expressed as a composition
\begin{equation}\label{omega_composite}
\omega=\tau\circ h:\,\omega(\Lambda)=\tau(h(\Lambda)),
\end{equation}
with $h$ in (\ref{H_map}) and
\begin{equation}\label{map_T_C}
\begin{split}
\tau\colon\Cscr_+ & \to\image_+\Gamma \\
C & \mapsto\int G\Phi_C G^*.
\end{split}
\end{equation}
Since $h$ has been proved to be a diffeomorphism, we can restrict our attention to the map $\tau$ due to the next simple result.
\begin{prop}
	Let $X,Y,Z$ be open subsets of $\Rbb^n$. Suppose we have functions $f:X\to Y$, $g:Y\to Z$ and $f$ is a diffeomorphism between $X$ and $Y$. Define the composite function
	\begin{equation}\label{h_composite}
	h=g\circ f:X\to Z.
	\end{equation}
	Then $h$ is a diffeomorphism between $X$ and $Z$ if and only if $g$ is a diffeomorphism between $Y$ and $Z$.
\end{prop}
\begin{pf}
	The ``if'' part is trivial since a composition of two diffeomorphisms is again a diffeomorphism. To see the converse, for $y\in Y$, let $x=f^{-1}(y)\in X$ and put it into (\ref{h_composite}) as an argument of $h$. Then one gets	
	\[g=h\circ f^{-1},\]
	which is again a composition of two diffeomorphisms.
\end{pf}

Since properness of the map $\omega$ has already be proven, it remains to show that $\omega$ is continuously differentiable and has everywhere nonvanishing Jacobian. In view of the relation (\ref{omega_composite}) and the previous proposition, it would be sufficient and necessary that the map $\tau$ possesses such two properties. We need the next lemma before proving the continuous differentiability.
\begin{lem}\label{lem_seq_Lambda_bounded}
	Let a sequence $\{\Lambda_k\}_{k\geq1}\subset\Lscr_+$ converge to some $\bar{\Lambda}\in\Lscr_+$. Then there exists a real number $\mu>0$ such that
	\[G^*(e^{i\theta})\Lambda_k G(e^{i\theta})\geq\mu I,\quad\forall k, \theta.\]
\end{lem}
\begin{pf}
	The claim of the lemma follows from the continuity of the function $G^*(e^{i\theta})\Lambda G(e^{i\theta})$ in $\Lambda$ and $\theta$, and the uniform convergence of the sequence of functions $\{G^*\Lambda_k G\}_{k\geq1}$ to $G^*\bar{\Lambda}G$.
\end{pf}

\begin{prop}
	The map $\tau$ in $(\ref{map_T_C})$ is of class $C^1$.
\end{prop}
\begin{pf}
	We can proceed by mimicking the proof of \cite[Lemma 1]{Zhu-Baggio-17}, although the argument here is slightly more general. First compute the differential of $\Phi(z;C)$ w.r.t. $C\in\Cscr_+$ as
	\begin{equation}\label{delta_Phi_C}
	\delta\Phi(z;C;\delta C)=-(CG)^{-1}\delta CG\Phi_C-\Phi_CG^*\delta C^*(CG)^{-*},
	\end{equation}
	which is easily seen to be continuous in $C$ and $\theta\in[-\pi,\pi]$ for a fixed $\delta C\in\Cfrak$. 
	This means that we can take the differential of the map $\tau$ inside the integral in (\ref{map_T_C})
	\begin{equation}\label{delta_T}
	\delta \tau(C;\delta C)=\int G\delta\Phi(e^{i\theta};C;\delta C)G^*.
	\end{equation}
	Next we show that the above differential is continuous in $C$ for a fixed $\delta C$. To this end, suppose we have a sequence $\{C_k\}_{k\geq1}\subset\Cscr_+$ that converges to some $\bar{C}\in\Cscr_+$ as $k\to\infty$. Due to the relation (\ref{spec_fact_Lambda}), we have for each $k$
	\begin{equation}\label{Lambda_k_Ck}
	G^*\Lambda_k G=G^*C_k^*C_kG, \quad \forall z\in\Tbb,
	\end{equation}
	where $\Lambda_k=h^{-1}(C_k)\in\Lscr_+^\Gamma$. Since $h$ is a diffeomorphism by Theorem \ref{thm_diffeo_H}, we have 
	\[\lim_{k\to\infty}\Lambda_k=\bar{\Lambda}:=h^{-1}(\bar{C}).\]
	Let $\lambda_{\min,k}(\theta)$ be the smallest eigenvalue of $G^*(e^{i\theta})\Lambda_k G(e^{i\theta})$, and $\sigma_{\min,k}(\theta)$ be the smallest singular value of $C_k G(e^{i\theta})$. In view of (\ref{Lambda_k_Ck}), we have 
	\[\lambda_{\min,k}(\theta)=\sigma^2_{\min,k}(\theta)\]
	By Lemma \ref{lem_seq_Lambda_bounded}, there exist a real number $\mu>0$ such that
	\[\lambda_{\min,k}(\theta)\geq\mu\implies \sigma_{\min,k}(\theta)\geq\sqrt{\mu}, \quad \forall k,\theta.\]
    Then we have
    \[\begin{split}
    \|\delta\Phi(e^{i\theta};C_k;\delta C)\|_2 & \leq 2\|(C_kG)^{-1}\delta CG\Phi_{C_k}\|_2 \\
     & \leq 2\|(C_kG)^{-1}\|_2^3\|\delta CG\|_2\|\Psi\|_2 \\
     & \leq \frac{2}{\sigma^3_{\min,k}(\theta)}\|\delta CG\|_F\|\Psi\|_F\leq K, \\
    \end{split}\]
	where the constant
	\[K=\frac{2}{\mu^{3/2}}\max_\theta\|\delta CG(e^{i\theta})\|_F\max_\theta\|\Psi(e^{i\theta})\|_F.\]
	We can now bound the integrand in (\ref{delta_T}). For any $\theta\in[-\pi,\pi]$ and $k\geq1$, we have
	\[\begin{split}
	\left|\left[G\delta\Phi(e^{i\theta};C_k;\delta C)G^*\right]_{j\ell}\right| & \leq\|G\delta\Phi(e^{i\theta};C_k;\delta C)G^*\|_F \\
	 & \leq \kappa\|G\delta\Phi(e^{i\theta};C_k;\delta C)G^*\|_2 \\
	 & \leq\kappa K\|G\|_2^2\leq\kappa KG_{\max},
	\end{split}\]
	where 
	\begin{equation}\label{G_max}
		G_{\max}:=\max_{\theta\in[-\pi,\pi]}\trace\left\{G(e^{i\theta})G^*(e^{i\theta})\right\}
	\end{equation}
	and $\kappa$ is a constant for norm equivalence. The last step is an application of Lebesgue's dominated convergence theorem to conclude
	\[\lim_{k\to\infty}\delta \tau(C_k;\delta C)=\delta \tau(\bar{C};\delta C),\]
	which completes the proof.
	
\end{pf}



We are now left with the task of investigating whether the Jacobian of $\tau$ vanishes nowhere, which can be approached via the differential (\ref{delta_T}).
However, the trick of orthogonality in the proof of Proposition \ref{prop_diffeo_scalar} does not apply in a straightforward manner to the general map $\omega$. The desired result can be obtained if an additional constraint is imposed on the prior $\Psi$, and this is reported in the next proposition.

\begin{prop}\label{prop_conditioned_prior}
If the prior $\Psi$ is such that the equality
\begin{equation}\label{equal_tr_int}
\trace\int F^*\Psi F=\trace\int F\Psi F^*
\end{equation}
holds for any $C\in\Cscr_+$ and any $V\in\Cfrak$, where the matrix function $F=VG(CG)^{-1}$, then the Jacobian determinant of $\tau$ vanishes nowhere in $\Cscr_+$, and hence the map $\omega$ is a diffeomorphism.
\end{prop}
\begin{pf}
Fix $C\in\Cscr_+$ and let $\delta \tau(C;V)=0$ for some $V\in\Cfrak$. In view of (\ref{delta_T}), this would imply that
\[\delta\Phi(z;C;V)\in\ker\Gamma=(\image\,\Gamma^*)^\perp,\]
which in view of (\ref{Gamma_star}), means
\begin{equation}\label{ortho_imGamma_star}
\begin{split}
\langle G^*XG,\delta\Phi(z;C;V)\rangle & =\trace\int G^*XG\,\delta\Phi(z;C;V) \\
& =0,\quad \forall X\in\Hfrak_n.
\end{split}
\end{equation}

Choosing $X=C^*V+V^*C$ in (\ref{ortho_imGamma_star}) would lead to the relation
\[
\trace\int2F\Psi F^*+\Psi F^*F^*+FF\Psi=0
\]
after some manipulations of the variables using (\ref{delta_Phi_C}). The left-hand side in the above equation is different from
\begin{equation}\label{tr_int_sym_Psi}
\trace\int\left(F+F^*\right)\Psi\left(F+F^*\right)
\end{equation}
in only one term. If the equality (\ref{equal_tr_int}) holds for any $C\in\Cscr_+$ and $V\in\Cfrak$, then we would have the expression (\ref{tr_int_sym_Psi}) equal to zero, which, by the same reasoning as in Proposition \ref{prop_diffeo_scalar}, implies
\[F+F^*\equiv0,\quad \forall z\in\Tbb,\] 
which is equivalent to (\ref{rational_eqn}). In view of Proposition \ref{prop_rational_mat_eqn}, this in turn implies $V=0$.

\end{pf}

The above proposition does not improve much over Proposition \ref{prop_diffeo_scalar} for the scalar case, since the requirement on the prior seems very artificial and a matrix-valued $\Psi$ in general does not satisfy it, as illustrated in the next example.

\begin{ex}
Consider a static case in which $n=m$, $B=I$, and the matrix $A$ void, that is, the transfer function (\ref{trans_func}) reduces to $G=z^{-1}I$ and the output of the linear system is identical to the 1-step delayed input. Let us fix
$\Psi\equiv\diag\{1,2\}$ and $C=I$, and then (\ref{equal_tr_int}) would reduce to
\[\trace\, V^*\Psi V=\trace\, V\Psi V^*.\]
The only requirement on $V$ is being lower-triangular with real diagonal entries. Hence we can take, e.g.,
\[V=\bmat1&0\\1&2\\\emat,\]
and it is straightforward to check that the above equality does not hold. However, in this overly simplified example, the solution to Problem \ref{spec_estimation} is still unique. Indeed, given $\Sigma\in\image_+\Gamma$ and $\Psi\in\Sscr_m$, one is looking for a parameter $C\in\Cscr_+$ such that
\[\int C^{-1}\Psi C^{-*}=\Sigma.\]
Clearly, this implies
\[C^{-1}L_R=L_{\Sigma}U,\]
where the notation $L_A$ denotes the usual Cholesky factor of $A>0$, $U$ is a unitary matrix, and $R:=\int\Psi>0$. It then follows that $U$ is lower triangular with real and positive diagonal entries, since such are all $C$, $L_R$, and $L_\Sigma$. Hence $U$ is necessarily equal to identity, and $C=L_RL_\Sigma^{-1}$. 
This means that the condition on the prior in Proposition \ref{prop_conditioned_prior} is not necessary for the uniqueness of the solution.
\end{ex}

\section{Conclusion}

We have shown that a parametric spectral estimation problem is well-posed if the chosen prior is special. It would be interesting to investigate whether the claim would still hold when the prior is arbitrarily matrix-valued, and this is left for future work.


\bibliography{ifacconf}             
\end{document}